\documentclass[preprint]{elsarticle}
\usepackage{pstricks}
\usepackage{amsfonts}
\usepackage{amsmath}
\usepackage{graphicx}
\usepackage{amssymb}

\journal{Wave Motion}

\begin{document}

\begin{frontmatter}
\title{Eigenfrequency correction of Bloch-Floquet waves in a thin periodic bi-material strip with cracks lying on perfect and imperfect interfaces}
\author[imaps]{A. Vellender\fnref{fnasv}}
\ead{asv09@aber.ac.uk}
\author[imaps]{G.S. Mishuris\corref{corggm}\fnref{fnggm}}
\ead{ggm@aber.ac.uk}
\address[imaps]{Institute of Mathematics and Physics, Aberystwyth University, Physical Sciences Building, Aberystwyth, Ceredigion, SY23 3BZ.}
\cortext[corggm]{Corresponding author.}
\fntext[fnasv]{AV would like to thank Aberystwyth University for providing APRS funding.}
\fntext[fnggm]{GM is grateful for support from the European Union Seventh Framework Programme under contract number PIAP-GA-2009-251475.}
\begin{abstract}
We analyse an asymptotic low-dimensional model of anti-plane shear in a thin bi-material strip containing a periodic array of interfacial cracks. Both ideal and non-ideal interfaces are considered. We find that the previously derived asymptotic models display a degree of inaccuracy in predicting standing wave eigenfrequencies and suggest an improvement to the asymptotic model to address this discrepancy. Computations demonstrate that the correction to the standing wave eigenfrequencies greatly improve the accuracy of the low-dimensional model.
\end{abstract}
\end{frontmatter}

\section{Introduction}
In this paper we present a method to correct discrepancies which arise in the asymptotic approximation of standing wave eigenfrequencies in a thin waveguide containing cracks and different types of interface.

Substantial interest in the analysis of waves interacting with waveguide boundaries can be found in the literature. In acoustics and water waves, problems in periodic waveguides have been studied in \cite{Evans,McIver,Linton}, among others. Bi-material structures are widely used across many engineering disciplines, ranging from film coatings to armour production. In such applications, it becomes vital to understand how waves propagate through such structures and to estimate the stresses near the crack tip which may be sufficient to cause defects to propagate, which in turn may cause failure of the entire structure \cite{Mikata}.

The effect that interfacial cracks and other defects have upon the behaviour of structures is a particularly active research area attracting significant attention. Modelling of interfacial cracks was studied in the important early papers \cite{Hutchinson,Willis}.  The study of different types of interfaces, including ideal and non-ideal, is also widely covered in the literature, for example in \cite{Benveniste,Bostrom,Golub,Leunpichcharoen,Lipton}, among others. Recently the interaction of an interfacial crack with small impurities has been considered in the asymptotic regime in \cite{PiccInclusions}.

The present paper builds upon the results of \cite{Mish2007} and \cite{Vellender2011} and makes a breakthrough in improving the results in a wide range of cases that are important for applications. The former manuscript uses a weight function approach to construct an asymptotic model for out-of-plane Bloch Floquet wave propagation in a thin bi-material strip with an array of cracks positioned along the join. The latter considers a similar geometry, but with non-ideal interfaces lying between the cracks. Both papers use a weight function approach to obtain constants describing stress distribution near the crack tips and also dervice junction conditions for an asymptotic low-dimensional model. The conditions enable dispersion diagrams to be constructed. Comparison with finite element simulations demonstrate that in both the perfect and imperfect interface cases, the low dimensional model has high accuracy in the cases of the waves that propagate through the strip, usually of the order of $10^{-4}\%$, but display a larger discrepancy in the case of the standing waves. The size of this discrepancy depends greatly upon material and geometrical parameters but is typically somewhere in the region of 3-15\%. It is this standing wave discrepancy which we aim to address in this paper, by considering the next asymptotic terms for the solution and eigenfrequencies.

The structure of the present paper is as follows. In Section \ref{section:formulation} we present the problem formulation before developing the low dimensional model in Section \ref{section:ldm}. The eigenvalue correction term is derived in Section \ref{section:omega1}, firstly for an elementary symmetric and homogeneous special case which is easily traceable and enables us to derive the correction term for the standing wave analytically, before considering the fully general case. We then present numerical results which demonstrate the effectiveness of the correction method in a variety of cases and comment on situations where the model fails to give useful information.

\section{Problem Formulation}\label{section:formulation}
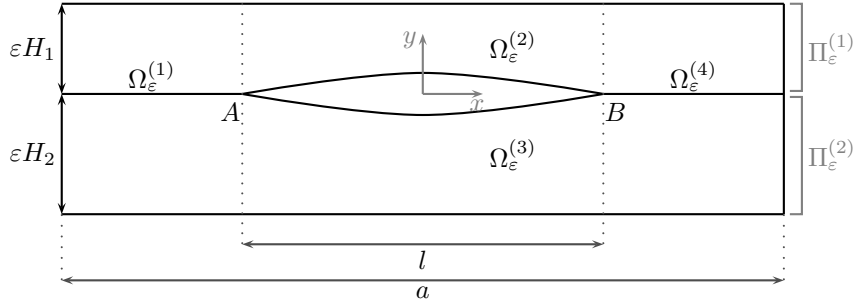
\begin{figure}[t]
\begin{center}
\psset{unit=0.8cm}
\begin{pspicture}[showgrid=false](-6,-3.3)(7,1.5)
\newgray{asv}{0.9}
\newgray{darker}{0.3}
\qline(-6,-2)(6,-2)
\qline(-6,1.5)(6,1.5)
\qline(-6,0)(-3,0)
\qline(3,0)(6,0)
\qline(6,1.5)(6,-2)
\psline{<->}(-6,-2)(-6,0)
\rput[rm](-6.1,-1){$\varepsilon H_2$}
\psline{<->}(-6,0)(-6,1.5)
\rput[rm](-6.1,0.75){$\varepsilon H_1$}
\pscurve(-3,0)(0,0.35)(3,0)
\pscurve(-3,0)(0,-0.35)(3,0)
\psline[linecolor=gray]{->}(0,0)(0,1)
\psline[linecolor=gray]{->}(0,0)(1,0)
\rput[tm](0.9,-0.1){\textcolor{gray}{$x$}}
\rput[rm](-0.1,0.9){\textcolor{gray}{$y$}}
\psline[linestyle=dotted,linecolor=darker](-3,-2.5)(-3,1.5)
\psline[linestyle=dotted,linecolor=darker](3,-2.5)(3,1.5)
\psline[linecolor=darker]{<->}(-3,-2.5)(3,-2.5)
\rput[tm](0,-2.6){$l$}
\psline[linestyle=dotted,linecolor=darker](-6,-3.1)(-6,-2)
\psline[linestyle=dotted,linecolor=darker](6,-3.1)(6,-2)
\psline[linecolor=darker]{<->}(-6,-3.1)(6,-3.1)
\rput[tm](0,-3.2){$a$}
\rput(-4.5,0.3){$\Omega^{(1)}_\varepsilon$}
\rput(1.5,0.8){$\Omega^{(2)}_\varepsilon$}
\rput(1.5,-1){$\Omega^{(3)}_\varepsilon$}
\rput(4.5,0.3){$\Omega^{(4)}_\varepsilon$}
\rput(-3.2,-0.3){$A$}
\rput(3.2,-0.3){$B$}
\psline[linecolor=gray](6.1,1.5)(6.3,1.5)(6.3,0.05)(6.1,0.05)
\psline[linecolor=gray](6.1,-0.05)(6.3,-0.05)(6.3,-2)(6.1,-2)
\rput[lm](6.4,0.75){\textcolor{gray}{$\Pi_\varepsilon^{(1)}$}}
\rput[lm](6.4,-1){\textcolor{gray}{$\Pi_\varepsilon^{(2)}$}}
\end{pspicture}
\end{center}
\caption{Geometry of the elementary cell.}
\label{cellgeom}
\end{figure}
The full problem formulation is as given in \cite{Vellender2011}, but is summarised here. The geometry considered is a bi-material strip composed of two materials of shear moduli $\mu_1$ and $\mu_2$, with respective thicknesses $\varepsilon H_1$ and $\varepsilon H_2$. The elementary cell of the periodic structure is shown in Figure \ref{cellgeom}. Along the interface lies a periodic array of cracks of length $l$, between which the interfaces are imperfect whose extent of imperfection is described by the parameter $\kappa$. The problem is singularly perturbed and so the case $\kappa=0$ which corresponds to the perfect interface case requires different analysis \cite{Mish2007} to the imperfect case \cite{Vellender2011}. The distance between adjacent cracks is $a-l$. The functions $u^{(j)}(x,y)$, $j=1,2,$ are respectively defined above and below the interface as solutions of the Helmholtz equations
\begin{equation}\label{helmholtz}
 \Delta u^{(j)}(x,y)+\frac{\omega^2}{c_j^2}u^{(j)}(x,y)=0.
\end{equation}
A zero stress component is imposed in the out-of-plane direction along the top and bottom of the strip, as well as along the face of the crack itself:
\begin{eqnarray}
 {\sigma_{yz}^{(1)}(x,\varepsilon H_1)=0,\qquad}{\sigma_{yz}^{(2)}(x,-\varepsilon H_2)=0,}\quad{x\in(-a/2,a/2),}\label{topbc}
 \\{\sigma_{yz}^{(1)}(x,0^+)=0,\qquad}{\sigma_{yz}^{(2)}(x,0^-)=0,}\quad{x\in(-l/2,l/2)}.
\end{eqnarray}
Outside the crack, along the boundary between $\Pi_\varepsilon^{(1)}$ and $\Pi_\varepsilon^{(2)}$, the interface is described by the condition
\begin{equation}\label{imperfectbc}
 u^{(1)}(x,0^+)-u^{(2)}(x,0^-)=\varepsilon\kappa\sigma_{yz}^{(1)}(x,0^+),\quad x\in(-a/2,-l/2)\cup(l/2,a/2).
\end{equation}
Note again that the case $\kappa=0$ corresponds to a perfect interface, while $\kappa>0$ represents imperfect interfaces.
We also assume continuity of tractions across the interface
\begin{equation}\label{middlebc}
  \sigma_{yz}^{(1)}(x,0^+)=\sigma_{yz}^{(2)}(x,0^-),\quad x\in(-a/2,-l/2)\cup(l/2,a/2).
\end{equation}
Solutions sought, $u^{(j)}$, represent Bloch-Floquet waves, so at the ends of our elementary cell $x=\pm a/2$ we have for $j=1,2$ the Bloch-Floquet conditions
\begin{eqnarray}
 u^{(j)}(-a/2,y)&=&e^{-iKa}u^{(j)}(a/2,y),\qquad y\in(-\varepsilon H_2,\varepsilon H_1),
  \\\sigma_{xz}^{(j)}(-a/2,y)&=&e^{-iKa}\sigma_{xz}^{(j)}(a/2,y),\qquad y\in(-\varepsilon H_2,\varepsilon H_1).
\end{eqnarray}

\subsection{Asymptotic Ansatz}
In both \cite{Mish2007} and \cite{Vellender2011}, eigenfunctions $u(x,y)$ are approximated in the form
\begin{eqnarray}\label{uansatzold}
u(x,y,\varepsilon)&=&\sum\limits_{k=0}^N{\varepsilon^k}
\left\{
  \sum\limits_{m=1}^4\chi_m
  \left(
    v_m^{(k)}(x)+\varepsilon^2 V_m^{(k)}(x,Y)
  \right)\right.\nonumber\\
&+&\left.\left(
W_A^{(k)}(X_A,Y)+W_B^{(k)}(X_B,Y)
\right)
\right\}
+R_N(x,y,\varepsilon),
\end{eqnarray}
with scaled co-ordinates $X_A,$ $X_B$ and $Y$ introduced in the vicinity of the left and right vertices of the crack defined as
\begin{equation}\label{scaledcoords}
 X_A=\frac{x-x_A}{\varepsilon},\qquad X_B=\frac{x-x_B}{\varepsilon},\qquad Y=\frac{y}{\varepsilon}.
\end{equation}
Here, $v_m^{(k)}$ represent solutions of lower-dimensional problems within limit sets $\Omega_0^{(j)}$, $j=1,2,3,4$. $\chi_m=\chi_m(x,y,\varepsilon)$ are cut-off functions defined so that $\chi_m(x,y;\varepsilon)\equiv1$ in $\Omega_\varepsilon^{(m)}$ and decay rapidly to zero outside $\Omega_\varepsilon^{(m)}$. They vanish near the so-called junction points $A$ and $B$ (the vertices of the crack). The terms $W_A^{(k)}$ and $W_B^{(k)}$ represent the boundary layers near $A$ and $B$, and  $V_m^{(k)}$ is the `fast' change of eigenfunctions in the transverse direction in the domain $\Omega_\varepsilon^{(j)}$. $R_N$ is the remainder term in the asymptotic approximation.

The asymptotic approach described is employed in both \cite{Mish2007} and \cite{Vellender2011}. When solving the low dimensional model to find the functions $v_m^{(k)}$, computations display a discrepancy for standing wave eigenfrequencies between the low order model and finite element simulations of the problem; these computations are presented in Section \ref{section:numerics}. In order to address the discrepancies that arise in this asymptotic model, in this paper we further consider the square of the frequency, $\omega^2$, as an asymptotic quantity, writing 
\begin{equation}\label{asymptoticrepresentationomega}
\omega^2=\sum\limits_{k=0}^N{\varepsilon^k\omega_k^2}.
\end{equation}
It is not immediately apparent {\em a priori} that this amendment will lead to a large correction in the approximations of eigenfrequencies, but we will later see that this allows us to solve the first order low-dimensional model which causes a significant improvement in the accuracy of the model in those cases where the zero order model displays large discrepancies. Interestingly, in cases where the zero order model gives high accuracy, the corrections are very small. As an example, in one case we consider in Section \ref{section:numerics}, the first-order correction method alters the frequency of the first standing wave (for which the zero-order model gives a significant discrepancy) by 11\% of its zero-order value, while the propagating waves (for which the zero-order model displays high accuracy) are only corrected by $10^{-4}\%$.

\section{Solution of low-dimensional model equations}\label{section:ldm}

Since the boundary layers $W_A$ and $W_B$ (see (\ref{uansatzold})) decay exponentially, we have that far from the crack tip, $\chi_m=1$,
\begin{equation}
 u\approx\sum_{k=0}^N\varepsilon^k\sum_{m=1}^4{\left(v_m^{(k)}+\varepsilon^2V_m^{(k)}\right)}.
\end{equation}
Substitution of this expression into the Helmholtz equation (\ref{helmholtz}) and comparing coefficients of terms in $\varepsilon^k$, $k=0,1$ respectively yields the two equations
\begin{equation}\label{zeroordercomparing}
 (v_m^{(0)})''+\frac{\partial^2 V_m^{(0)}}{\partial Y^2}+\frac{\omega_0^2}{c_j^2}v_m^{(0)}=0,\quad m=1,2,3,4;\quad j=1,2,
\end{equation}
\begin{equation}\label{firstordercomparing}
 (v_m^{(1)})''+\frac{\partial^2 V_m^{(1)}}{\partial Y^2}+\frac{\omega_1^2}{c_j^2}v_m^{(0)}+\frac{\omega_0^2}{c_j^2}v_m^{(1)}=0, \quad m=1,2,3,4;\quad j=1,2.
\end{equation}
Expression (\ref{zeroordercomparing}) corresponding to terms in $k=0$ is the same as before when $\omega$ was not treated as an asymptotic series, but (\ref{firstordercomparing}) is new.
Above and below the crack, that is for $m=2,3,$ we therefore have that
\begin{equation}\label{crackd2d3}
 \frac{\partial^2 V_m^{(1)}}{\partial Y^2}=-\left[(v_m^{(1)})''+\frac{\omega_1^2}{d_m^2}v_m^{(0)}+\frac{\omega_0^2}{d_m^2}v_m^{(1)}\right], \quad m=2,3,
\end{equation}
where $d_2=c_1$ and $d_3=c_2$, which after integration and application of the boundary condition $\left.\frac{\partial V_m^{(1)}}{\partial Y}\right|_{Y=0\pm}\equiv0$ yields the equation
\begin{equation}\label{firstorder23}
 (v_m^{(1)})''(x)+\frac{\omega_0^2}{d_m^2}v_m^{(1)}(x)+\frac{\omega_1^2}{d_m^2}v_m^{(0)}(x)=0,\quad m=2,3.
\end{equation}
For $m=1,4$ (where no crack is present), rearranging and integrating (\ref{firstordercomparing}) and applying the condition for continuity of tractions across the imperfect interface yields the equation
\begin{equation}\label{nocrackd1}
 (v_m^{(1)})''(x)+\frac{\omega_0^2}{d_1^2}v_m^{(1)}(x)+\frac{\omega_1^2}{d_1^2}v_m^{(0)}(x)=0,
\end{equation}
where
\begin{equation}
 d_1=c_1c_2\sqrt{\frac{\mu_1H_1+\mu_2H_2}{\mu_1H_1c_2^2+\mu_2H_2c_1^2}}.
\end{equation}
For the zero order approximation,
\begin{equation}\label{ode1}
 (v_m^{(0)})''(x)+\frac{\omega_0^2}{d_m^2}v_m^{(0)}(x)=0,\quad m=1,2,3,4.
\end{equation}

\subsection{Junction conditions and crack tip asymptotics}
The asymptotic representation of $\omega$ does not affect junction conditions on the first two levels of the approximation. Junction conditions for the zero order approximation have been derived in \cite{Vellender2011} and read
\begin{equation}\label{zerojcn1}
 v_1^{(0)}(x_A)=v_2^{(0)}(x_A)=v_3^{(0)}(x_A);\quad
 v_2^{(0)}(x_B)=v_3^{(0)}(x_B)=v_4^{(0)}(x_B),
\end{equation}
along with the conditions for flux 
\begin{equation}\label{zerojcn3}
 \mu_1H_1(v_2^{(0)})'(x_A)+\mu_2H_2(v_3^{(0)})'(x_A)=(\mu_1H_1+\mu_2H_2)(v_1^{(0)})'(x_A).
\end{equation}
\begin{equation}\label{zerojcn4}
 \mu_1H_1(v_2^{(0)})'(x_B)+\mu_2H_2(v_3^{(0)})'(x_B)=(\mu_1H_1+\mu_2H_2)(v_4^{(0)})'(x_B).
\end{equation}
Junction conditions for the first order approximation at the right hand crack tip are given for $m=2,3,$ by
\begin{equation}\label{jncondfirst}
 v_m^{(1)}(x_B)=v_4^{(1)}(x_B)+(-1)^{m+1}\frac{\mu_2H_2}{\mu_1H_1+\mu_2H_2}\alpha_N\Delta\{(v^{(0)})'\}(x_B),
\end{equation}
where $N=P$ if $\kappa=0$ (the perfect interface case) and $N=I$ if $\kappa>0$ (the imperfect interface case), and
\begin{equation}
 \Delta\{(v^{(0)})'\}(x)=(v_2^{(0)})'(x)-(v_3^{(0)})'(x).
\end{equation}
 We will continue to use this notation throughout the rest of the manuscript. The junction conditions (\ref{jncondfirst}) are valid for both perfect and imperfect cases, but the form of the corresponding constants $\alpha_P$ and $\alpha_I$ are absolutely different and come from different analysis; this arises from the fact that the problem is singularly perturbated and so different analysis is needed in the cases $\kappa>0$ to the case $\kappa=0$. The definitions of the constants $\alpha_I$ and $\alpha_P$ are derived in \cite{Vellender2011} and \cite{Mish2007} respectively and stated here. For the perfect interface case, the constant is defined as
\begin{equation}
 \alpha_P=\frac{H_1+H_2}{\pi}\left\{\mu_*\int\limits_0^\infty{f(t)}dt-\ln\left\{\left(\frac{1+H_*}{2}\right)^{\frac{1+H_*}{2}}\left(\frac{1-H_*}{2}\right)^{\frac{1-H_*}{2}}\right\}\right\},
\end{equation}
where
\begin{equation}
 f(t)=\frac{H_*-\tanh(tH_*)\coth(t)}{(\sinh(t)+\mu_*\sinh(tH_*))t},\quad\mu_*=\frac{\mu_1-\mu_2}{\mu_1+\mu_2},\quad H_*=\frac{H_1-H_2}{H_1+H_2}.
\end{equation}
In the imperfect interface case, the constant is given by
\begin{equation}
 \alpha_I=(H_1+H_2)\left\{\frac{1}{\pi}\int\limits_0^\infty{\frac{\ln g(t)}{t^2}}dt+\frac{1}{\lambda_*}\right\},
\end{equation}
where 
\begin{equation}
 g(t)=\frac{t}{\lambda_*^2+t^2}\left(t+\frac{2}{\kappa_*(1+\mu_*)}\coth\frac{t(1+H_*)}{2}+\frac{2}{\kappa_*(1-\mu_*)}\coth\frac{t(1-H_*)}{2}\right),
\end{equation}
\begin{equation}
\lambda_*=(H_1+H_2)\sqrt{\frac{\mu_1H_1+\mu_2H_2}{\mu_1\mu_2H_1H_2\kappa}},\quad\kappa_*=\frac{\kappa(\mu_1+\mu_2)}{(H_1+H_2)}.
\end{equation}
We stress that $\alpha_I$ is a constant that depends heavily upon $\kappa$ and so describes how the junction conditions are impacted by the imperfect interface. The first order fluxes satisfy the relationship
\begin{equation}\label{firstfluxes}
 (\mu_1H_1+\mu_2H_2)(v_4^{(1)})'(x_B)-\mu_1H_1(v_2^{(1)})'(x_B)-\mu_2H_2(v_3^{(1)})'(x_B)=0.
\end{equation}
The analogous conditions for $m=2,3,$ at the other vertex can be obtained by replacing $B$ by $A$ and $m+1$ by $m$ in equation (\ref{jncondfirst}). The other crack tip's condition for fluxes is as in equation (\ref{firstfluxes}), but again replacing $B$ by $A$.

The zero-order and first-order constants describing the singular behaviour of the full solution near the crack tips derived for the perfect and imperfect interface cases respectively in \cite{Mish2007} and \cite{Vellender2011} are unaffected by the consideration of $\omega$ as an asymptotic series. However, if one continues to deeper levels of the asymptotics, the junction conditions of fifth order and higher would be affected by taking $\omega$ as an asymptotic series.

\subsection{Corrected low dimensional model}
\subsubsection{Zero order low dimensional model}
Solutions of the zero order equation (\ref{ode1}) for $m=1,2,3,4$ are of the form
\begin{equation}\label{zeroorderform}
 v_m^{(0)}(x)=A_m^{(0)}\sin\left(\frac{\omega_0}{d_m} x\right)+B_m^{(0)}\cos\left(\frac{\omega_0}{d_m} x\right),
\end{equation}
The first order equation (\ref{ode1}) has solutions in the form
\begin{equation}\label{firstorderform}
 v_m^{(1)}(x)=A_m^{(1)}\sin\left(\frac{\omega_0}{d_m} x\right)+B_m^{(1)}\cos\left(\frac{\omega_0}{d_m} x\right)+\omega_1^2F_m(x),
\end{equation}
where
\begin{equation}
 F_m(x)=\frac{x}{2d_m\omega_0}\left\{A_m^{(0)}\cos\left(\frac{\omega_0}{d_m} x\right)-B_m^{(0)}\sin\left(\frac{\omega_0}{d_m} x\right)\right\}.
\end{equation}
We note that assuming the zero order system has been solved, all constants in this expression for $F_m(x)$ are considered known.

Let us first consider the zero order case. We see from (\ref{zeroorderform}) that eight constants need to be evaluated, $A_m^{(0)}$ and $B_m^{(0)}$ for $m=1,2,3,4$ which we write in the column vector $A^{(0)}$ defining the notation
\begin{equation}\label{columnAk}
 A^{(k)}=\left[
\begin{array}{c c c c c c c c}
   A_1^{(k)} & B_1^{(k)} & A_2^{(k)} & B_2^{(k)} & A_3^{(k)} & B_3^{(k)} & A_4^{(k)} & B_4^{(k)}
\end{array}
\right]^T.
\end{equation} 
We have six junction conditions: two from (\ref{zerojcn1}) and one from each of (\ref{zerojcn3}) and (\ref{zerojcn4}), which can be complemented by the two Bloch-Floquet conditions to yield the $8\times8$ matrix equation
\begin{equation}
 M_{8\times8}A^{(0)}=0.
\end{equation}
We introduce notation to abbreviate the entries of $M_{8\times8}$ as follows:
\begin{equation}
 S_m=\sin\left(\frac{d_1}{d_m}\varpi_0x_B\right);\quad C_m=\sin\left(\frac{d_1}{d_m}\varpi_0x_B\right);\quad m=1,2,3,4;
\end{equation}
\begin{equation}
  \psi_j=\frac{\mu_jH_j}{\mu_1H_1+\mu_2H_2}\frac{d_1}{d_{j+1}};\; j=1,2;\; S_a=\sin\left(\varpi_0\frac{a}{2}\right); C_a=\cos\left(\varpi_0\frac{a}{2}\right),
\end{equation}
where $\varpi_j=\omega_j/d_1$ is introduced for normalisation.
Now, $M_{8\times8}=$
\begin{equation}\label{M8x8}
\left[
  \begin{array}{c c c c c c c c}
    0&0&S_2&C_2&0&0&-S_4&-C_4\\
    0&0&0&0&S_3&C_3&-S_4&-C_4\\
    0&0&\psi_1C_2&-\psi_1S_2&\psi_2C_3&-\psi_2S_3&-C_4&S_4\\
    -S_1&-C_1&S_2&C_2&0&0&0&0\\
    S_1&-C_1&0&0&-S_3&C_3&0&0\\
    -C_1&-S_1&\psi_1C_2&\psi_1S_2&\psi_2C_3&\psi_2S_3&0&0\\
    -S_a&C_a&0&0&0&0&-e^{-iKa}S_a&-e^{-iKa}C_a\\
    C_a&S_a&0&0&0&0&-e^{-iKa}C_a&e^{-iKa}S_a
  \end{array}
\right]
\end{equation}
%
The determinant of $M_{8\times8}$ can be written in the form
\begin{equation}
 \det(M_{8\times8})=\mathcal{A}(\omega_0)e^{-2iKa}+\mathcal{B}(\omega_0)e^{-iKa}+\mathcal{A}(\omega_0).
\end{equation}
It can be shown that for the case in which all wavespeeds $d_m$ are equal to $d$, say, both $\mathcal{A}(\omega_0)$ and $\mathcal{B}(\omega_0)$ are zero when $\omega_0=n\pi d/(2x_B)$, $n\in\mathbb{N}$. It follows that in this case, the standing waves have no dependence upon the Bloch-Floquet parameter $K$. This observation motivates us to consider the special case discussed in subsection \ref{section:homsym}.

\subsubsection{First order low dimensional model}
Applying the junction and Bloch-Floquet conditions for the first order equation (\ref{ode1}) yields the matrix equation
\begin{equation}
 M_{8\times8}A^{(1)}=\varpi_1^2N_{8\times8}A^{(0)}+B_A\Delta\{(v^{(0)})'\}(x_A)+B_B\Delta\{(v^{(0)})'\}(x_B).
\end{equation}
Here, $M_{8\times8}$ is the matrix defined in (\ref{M8x8}), $A^{(0)}$ and $A^{(1)}$ are the coefficients defined in (\ref{columnAk}). Since $M_{8\times8}$, is singular, this equation gives a solvability condition which will allow us to find the correction term, $\omega_1$. The matrix $N_{8\times8}$ is defined as
\begin{equation}\label{N8x8}
\left[
  \begin{array}{c c c c c c c c}
    0&0&\frac{-d_1x_BC_2}{2d_2\varpi_0}&\frac{d_1x_BS_2}{2d_2\varpi_0}&0&0&\frac{x_BC_4}{2\varpi_0}&\frac{-x_BS_4}{2\varpi_0}\\
    0&0&0&0&\frac{-d_1x_BC_3}{2d_3\varpi_0}&\frac{d_1x_BS_3}{2d_3\varpi_0}&\frac{x_BC_4}{2\varpi_0}&\frac{-x_BS_4}{2\varpi_0}\\
    0&0&N_{3,3}&N_{3,4}&N_{3,5}&N_{3,6}&N_{3,7}&N_{3,8}\\
    \frac{-x_BC_1}{2\varpi_0}&\frac{-x_BS_1}{2\varpi_0}&\frac{d_1x_BC_2}{2d_2\varpi_0}&\frac{d_1x_BS_2}{2d_2\varpi_0}&0&0&0&0\\
    \frac{-x_BC_1}{2\varpi_0}&\frac{-x_BS_1}{2\varpi_0}&0&0&\frac{d_1x_BC_3}{2d_3\varpi_0}&\frac{d_1x_BS_3}{2d_3\varpi_0}&0&0\\
    N_{6,1}&N_{6,2}&N_{6,3}&N_{6,4}&N_{6,5}&N_{6,6}&0&0\\
    \frac{aC_a}{4\varpi_0}&\frac{aS_a}{4\varpi_0}&0&0&0&0&\frac{aZC_a}{4\varpi_0}&\frac{-aZS_a}{4\varpi_0}\\
    N_{8,1}&N_{8,2}&0&0&0&0&N_{8,7}&N_{8,8}
\end{array}
\right]
\end{equation}
where $Z=e^{-iKa}$ and for $q\in\{3,4,5,6,7,8\}$,
\begin{eqnarray}
 N_{3,q}&=&\frac{d_1}{d_{\lfloor\frac{q+1}{2}\rfloor}}x_B\left(\left(\frac{1+(-1)^q}{2}\right)C_{\lfloor\frac{q+1}{2}\rfloor}+\left(\frac{1-(-1)^q}{2}\right)S_{\lfloor\frac{q+1}{2}\rfloor}\right)\nonumber
\\&-&\frac{1}{\varpi_0}\left(\left(\frac{1-(-1)^q}{2}\right)C_{\lfloor\frac{q+1}{2}\rfloor}-\left(\frac{1+(-1)^q}{2}\right)S_{\lfloor\frac{q+1}{2}\rfloor}\right).
\end{eqnarray}
For $q\in\{1,2,3,4,5,6\}$,
\begin{eqnarray}
 N_{6,q}&=&\frac{(-1)^{q+1}d_1}{d_{\lfloor\frac{q+1}{2}\rfloor}}x_B\left(\left(\frac{1+(-1)^q}{2}\right)C_{\lfloor\frac{q+1}{2}\rfloor}+\left(\frac{1-(-1)^q}{2}\right)S_{\lfloor\frac{q+1}{2}\rfloor}\right)\nonumber
\\&+&\frac{(-1)^q}{\varpi_0}\left(\left(\frac{1-(-1)^q}{2}\right)C_{\lfloor\frac{q+1}{2}\rfloor}-\left(\frac{1+(-1)^q}{2}\right)S_{\lfloor\frac{q+1}{2}\rfloor}\right),
\end{eqnarray}
and the expressions of the eighth row are given by
\begin{equation}
 N_{8,1}=\frac{a}{4}S_a-\frac{1}{2\varpi_0}C_a,\quad N_{8,2}=\frac{1}{2\varpi_0}S_a-\frac{a}{4}C_a,
\end{equation}
\begin{equation}
 N_{8,7}=e^{-iKa}\left(\frac{1}{2\varpi_0}C_a-\frac{a}{4}S_a\right),\quad N_{8,8}=-e^{-iKa}\left(\frac{1}{2\varpi_0}S_a+\frac{a}{4}C_a\right).
\end{equation}
The vectors $B_A$ and $B_B$ are given by
\begin{equation}
 B_A=\alpha_N\left[
\begin{array}{c c c c c c c c}
   0&0&0-\frac{\mu_2H_2}{\mu_1H_1+\mu_2H_2}&\frac{\mu_1H_1}{\mu_1H_1+\mu_2H_2}&0&0&0
\end{array}
\right]^T,
\end{equation}
\begin{equation}
 B_B=\alpha_N\left[
\begin{array}{c c c c c c c c}
   -\frac{\mu_2H_2}{\mu_1H_1+\mu_2H_2}&\frac{\mu_1H_1}{\mu_1H_1+\mu_2H_2}&0&0&0&0&0&0
\end{array}
\right]^T.
\end{equation}

\section{Derivation of first order correction term, $\omega_1$}\label{section:omega1}
\subsection{Homogeneous Symmetric Case}\label{section:homsym}

In this section, we condsider the symmetric case in which $H_1=H_2$ and $\mu_1=\mu_2$. This simple case is instructive since the symmetry enables us to analytically determine the eigenfrequency of the first standing wave in an easily traceable process; we will later make indications on how the method for the general case relates to and differs from this procedure. Moreover, this eigenfrequency does not have any $K$-dependence as is the case for inhomogeneous setups, which enables us to easily separate the first standing wave solution from the others. In the case of the standing wave, the beams above and below the crack vibrate while the others do not (that is, $v_1^{(k)}(x)=v_4^{(k)}(x)=0$ for $k=0,1$).

The problem formulation for the symmetric case is as follows. For the zero-order approximation, solutions satisfy (\ref{ode1}) with $d_m=d$ for all $m=1,2,3,4$. To isolate the standing waves (whose frequencies we wish to impose a correction upon) we impose the condition
\begin{equation}
 v_1^{(0)}(x)\equiv v_4^{(0)}(x)\equiv0.
\end{equation}
The junction conditions for the zero order approximation then simplify to
\begin{equation}\label{symjunction1}
 v_2^{(0)}(x_A)=v_2^{(0)}(x_B)=0,\quad v_3^{(0)}(x_A)=v_3^{(0)}(x_B)=0,
\end{equation}
along with
\begin{equation}\label{symjunction4}
 (v_2^{(0)})'(x_A)+(v_3^{(0)})'(x_A)=0,\quad (v_2^{(0)})'(x_B)+(v_3^{(0)})'(x_B)=0.
\end{equation}
For $m=2,3,$ we have that the general solution of of the zero order LDM is of the form (\ref{zeroorderform}) with $d_m=d$. Applying conditions (\ref{symjunction1})-(\ref{symjunction4}) yields
\begin{equation}
 v_2^{(0)}(x)=B\cos\left(\frac{\omega_0}{d}x\right),\quad
 v_3^{(0)}(x)=-B\cos\left(\frac{\omega_0}{d}x\right).
\end{equation}

The first order approximation equation is of the form (\ref{firstorder23}) with $d_m=d$.
Since $v_m^{(0)}$ are now known functions, the corresponding system consists of the two ordinary differential equations
\begin{equation}
 (v_2^{(1)})''(x)+\frac{\omega_0^2}{d^2}v_2^{(1)}(x)+\frac{\omega_1^2}{d^2}B\cos\left(\frac{\omega_0}{d}x\right)=0,
\end{equation}
\begin{equation}
 (v_3^{(1)})''(x)+\frac{\omega_0^2}{d^2}v_3^{(1)}(x)-\frac{\omega_1^2}{d^2}B\cos\left(\frac{\omega_0}{d}x\right)=0.
\end{equation}
These ODEs have respective elementary solutions
\begin{equation}\label{v21form}
 v_2^{(1)}(x)=A_2^{(1)}\sin\left(\frac{\omega_0}{d}x\right)+B_2^{(1)}\cos\left(\frac{\omega_0}{d}x\right)-\omega_1^2\frac{Bx\sin\left(\frac{\omega_0}{d}x\right)}{2\omega_0 d},
\end{equation}
\begin{equation}\label{v31form}
 v_3^{(1)}(x)=A_3^{(1)}\sin\left(\frac{\omega_0}{d}x\right)+B_3^{(1)}\cos\left(\frac{\omega_0}{d}x\right)+\omega_1^2\frac{Bx\sin\left(\frac{\omega_0}{d}x\right)}{2\omega_0 d},
\end{equation}
complemented by the juction conditions (which follow from \cite{Vellender2011})
\begin{equation}\label{symjunction5}
 v_m^{(1)}(x_\beta)=\frac{(-1)^{m+p}}{2}\alpha_N\Delta\left\{(v^{(0)})'\right\}(x_\beta),\quad m=2,3;\; \beta=A,B,
\end{equation}
where $p=0$ if $\beta=A$ and $p=1$ if $\beta=B$.
\begin{figure}[t]
\begin{center}
 \includegraphics[width=0.5\linewidth]{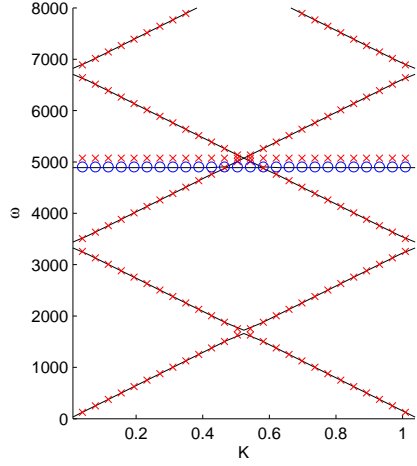}
\end{center}
  \caption{{\footnotesize{Dispersion diagram for the perfect interface case with medium crack length $l=2$m in an elementary cell with $a=6$m. The material above and below the crack is iron. The solid black lines show finite element results, while red crosses ({\textcolor{red}{$\times$}}) show the zero order approximation and blue circles (\textcolor{blue}{$\circ$}) show the corrected regime, with the first standing wave corrected by the analytic derivation of $\omega_1$ as presented in equation (\ref{analyticomega}).}}}
  \label{analyticfigure}
\end{figure}
For the first standing wave, $\omega_0={\pi d}/{l}$, and so applying (\ref{symjunction5}) with $\beta=B$, $m=2$, we see that
\begin{equation}
 A_2^{(1)}=\frac{B}{d^2}\left(\omega_1^2\frac{l^2}{4\pi}+\omega_0\alpha_N\right).
\end{equation}
Applying the condition (\ref{symjunction5}) with $\beta=A$, $m=2$, yields the condition
\begin{equation}
 A_2^{(1)}=-\frac{B}{d^2}\left(\omega_1^2\frac{l^2}{4\pi}-\omega_0\alpha_N\right).
\end{equation}
Since $A_2^{(1)}$ is equal to the positive and negative of the same function, it must be zero, from which we deduce that
\begin{equation}\label{analyticomega}
 \omega_1^2=-\frac{4\pi\omega_0\alpha_N}{l^2},
\end{equation}
Thus, for the symmetrical case where both materials are the same, we have found an expression for the correction term $\omega_1^2$.

Figure \ref{analyticfigure} shows the dispersion diagram for this homogeneous, symmetric case. The red crosses on the diagram indicate the zero-order approximation ($\omega_0$) of the eigenfrequencies, while the blue circles show the corrected first-order approximation. The black lines result from a finite element computation. The derived correction method improves the standing wave frequency discrepancy from $3.7\%$ to just $0.26\%$.

\subsection{General case}
%
%
%

We now consider the general case in which the materials above and below the crack and interface may have different thicknesses and shear moduli.
For the first order approximation, after the application of junction and Bloch-Floquet conditions we obtain a matrix equation of the form
\begin{equation}\label{1stmatrixeqn}
 MA^{(1)}=NA^{(0)}+B_A\Delta\{(v^{(0)})'\}(x_A)+B_B\Delta\{(v^{(0)})'\}(x_B).
\end{equation}
Here, $M$ and $N$ are both $8\times8$ matrices as defined earlier in (\ref{M8x8}) and (\ref{N8x8}) respectively whose elements depend on the Bloch-Floquet parameter $K$ and the eigenfrequency $\omega_0$ which is such that $\det(M)=0$. Because $\det(M)=0$, $M$ has zero among its eigenvalues. We can write
\begin{equation}\label{diagM}
 M=VDV^{-1}
\end{equation}
where V is a matrix whose columns are eigenvectors of $M$ and $D$ is a diagonal matrix with the respective eigenvalues of $M$ along the diagonal. Premultiplying (\ref{1stmatrixeqn}) by $V^{-1}$, we can write
\begin{equation}
  V^{-1}MVV^{-1}A^{(1)}=V^{-1}(\omega_1^2NA^{(0)}+B_A\Delta\{(v^{(0)})'\}(x_A)+B_B\Delta\{(v^{(0)})'\}(x_B)),
\end{equation}
which upon substitution of (\ref{diagM}) becomes
\begin{equation}\label{1stmatrixeqnwithD}
 DV^{-1}A^{(1)}=\omega_1^2V^{-1}NA^{(0)}+V^{-1}B_A\Delta\{(v^{(0)})'\}(x_A)+V^{-1}B_B\Delta\{(v^{(0)})'\}(x_B).
\end{equation}
Since $M$ is singular, it posesses zero as an eigenvalue, and so one row of the left hand side matrix in (\ref{1stmatrixeqnwithD}) is equal to zero. Let us denote that row $l$. Then
\begin{equation}
 \omega_1^2\left(V^{-1}NA^{(0)}\right)_l+\left(V^{-1}B_A\right)_l\Delta\{(v^{(0)})'\}(x_A)+\left(V^{-1}B_B\right)_l\Delta\{(v^{(0)})'\}(x_B)=0.
\end{equation}
All matrices and parameters in this equation are now known, with the exception of $\omega_1^2$ which can now be written in terms of known quantities:
\begin{equation}
 \omega_1^2=-\frac{\left(V^{-1}(B_A\Delta\{(v^{(0)})'\}(x_A)+B_B\Delta\{(v^{(0)})'\}(x_B))\right)_l}{\left(V^{-1}NA^{(0)}\right)_l}.
\end{equation}

%
A potential problem with this computational method is that the matrix V may have a determinant which is close to zero when eigenvalues of $M$ are close together. To eliminate any possible errors arising from this, we introduce a second computational scheme for computing $\omega_1$.

The Schur decomposition states that if $A$ is a $n\times n$ square matrix with complex entries, then $A$ can be expressed in the form $A=QUQ^{-1}$ where $Q$ is unitary and $U$ is upper triangular, with the eigenvalues of $A$. In our case, we apply Schur decomposition to the transpose of $M$:
\begin{equation}
 M^T=QUQ^{-1}.
\end{equation}
Schur decomposition is not unique; we may place the smallest eigenvalue in the first position along the leading diagonal, and since $M$ is singular, this eigenvalue is zero. The first column of the upper triangular matrix $U$ is therefore a row of zeros. Since $M=(Q^{-1})^TU^TQ^T$, we can premultiply the first order matrix equation by $Q^T$ and substutite to obtain
\begin{equation}
 U^TQ^TA^{(1)}=\omega_1^2Q^TNA^{(0)}+Q^TB_A\Delta\{(v^{(0)})'\}(x_A)+Q^TB_B\Delta\{(v^{(0)})'\}(x_B).
\end{equation}
The first row of the left hand side is zero, whence
\begin{equation}
 \omega_1^2=-\frac{\left(Q^T(B_A\Delta\{(v^{(0)})'\}(x_A)+B_B\Delta\{(v^{(0)})'\}(x_B))\right)_1}{\left(Q^TNA^{(0)}\right)_1}.
\end{equation}

\section{Numerical results}\label{section:numerics}
\subsection{Materials and geometries used in numerical simulations}

For our numerical calculations we will consider a strip whose elementary cell is of length $a=6$m with an overall thickness of $H_1+H_2=0.15$m. This geometry corresponds to a value of $\varepsilon=0.025$. We will compare results from the low dimensional model against those from finite element simulations (COMSOL). We stress that finite element simulatons are efficient for comparison only in cases when the strip is not too thin, i.e. when $\varepsilon$ is not too small. The low dimensional model, however, remains valid as $\varepsilon\to0$. 
For our computations we vary four parameters as listed in Appendix \ref{parametersappendix}. These parameters are type of interface (perfect, imperfect, highly imperfect), length of crack (short, medium, long), materials (iron/aluminium [similar wavespeeds], magnesium/aluminium [less similar wavespeeds]) and thicknesses of each material (symmetric geometry, asymmetric geometry).

We present in this section a number of dispersion diagrams, plotting frequency $\omega$ against the Bloch-Floquet parameter $K$. We refer to plots of $\omega=\omega_0$ as the zero order approximation, and to plots of $\omega$ calculated according to (\ref{asymptoticrepresentationomega}) as the first order approximation, or the corrected frequency.

\subsection{Perfect interface}
\begin{figure}[t]
  \includegraphics[width=0.9\linewidth]{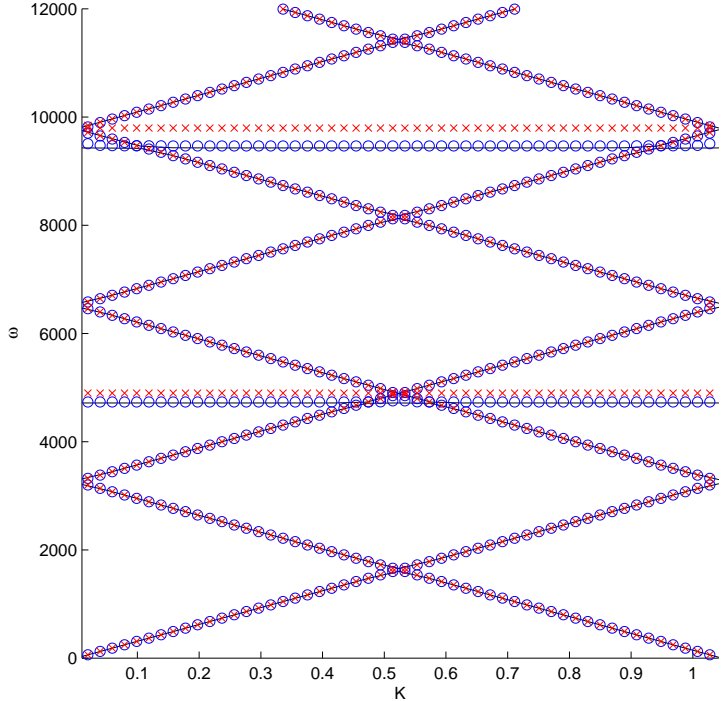}
  \caption{{\footnotesize{Dispersion diagram for a perfect-interface strip composed of equal thicknesses of aluminium and magnesium with a medium length crack ($l=2$m). The solid black lines show the finite element results, while red crosses ({\textcolor{red}{$\times$}}) show the zero order approximation and blue circles (\textcolor{blue}{$\circ$}) show the corrected first order approximation.}}}
  \label{homogeneous_correction}
\end{figure}

\subsubsection{Materials with similar wave speeds}
Figure \ref{homogeneous_correction} demonstrates the effectiveness of the method of eigenfrequency correction for the standing waves. This dispersion diagram results from computations corresponding to the case of a strip with a `sensible' crack length, composed of equal thicknesses of magnesium and aluminium (materials possessing similar wave speeds). The diagram demonstrates that the zero order approximation agrees to a very high degree of accuracy with the finite element results in the cases of the waves which propagate through the strip (the slanted lines). Typically the zero order approximation for these waves' eigenfrequencies differs from the finite element simulation only by around $10^{-4}\%$. However, it is clear that there is a discrepancy between the zero order model and the finite element results in the case of the standing waves (horizontal lines on the dispersion diagram). The corrected first-order model retains the excellent accuracy for propagating waves, slightly increasing the accuracy while remaining of the order of $10^{-4}\%$, and hugely improves the discrepancy of the standing wave frequencies. While the correction is not completely uniform since the standing waves' frequencies depend upon $K$ except in the case where materials have {\em identical} wavespeeds, a typical discrepancy for the first standing wave has decreased from 3.9\% to 0.38\%. This can be considered as a surprisingly large correction, since there is no reason {\em a priori} to suspect that considering $\omega$ as an asymptotic quantity should cause such an improvement in the accuracy of the low dimensional model's approximation of the standing wave frequency.

\begin{figure}[t]
 \includegraphics[width=0.45\linewidth]{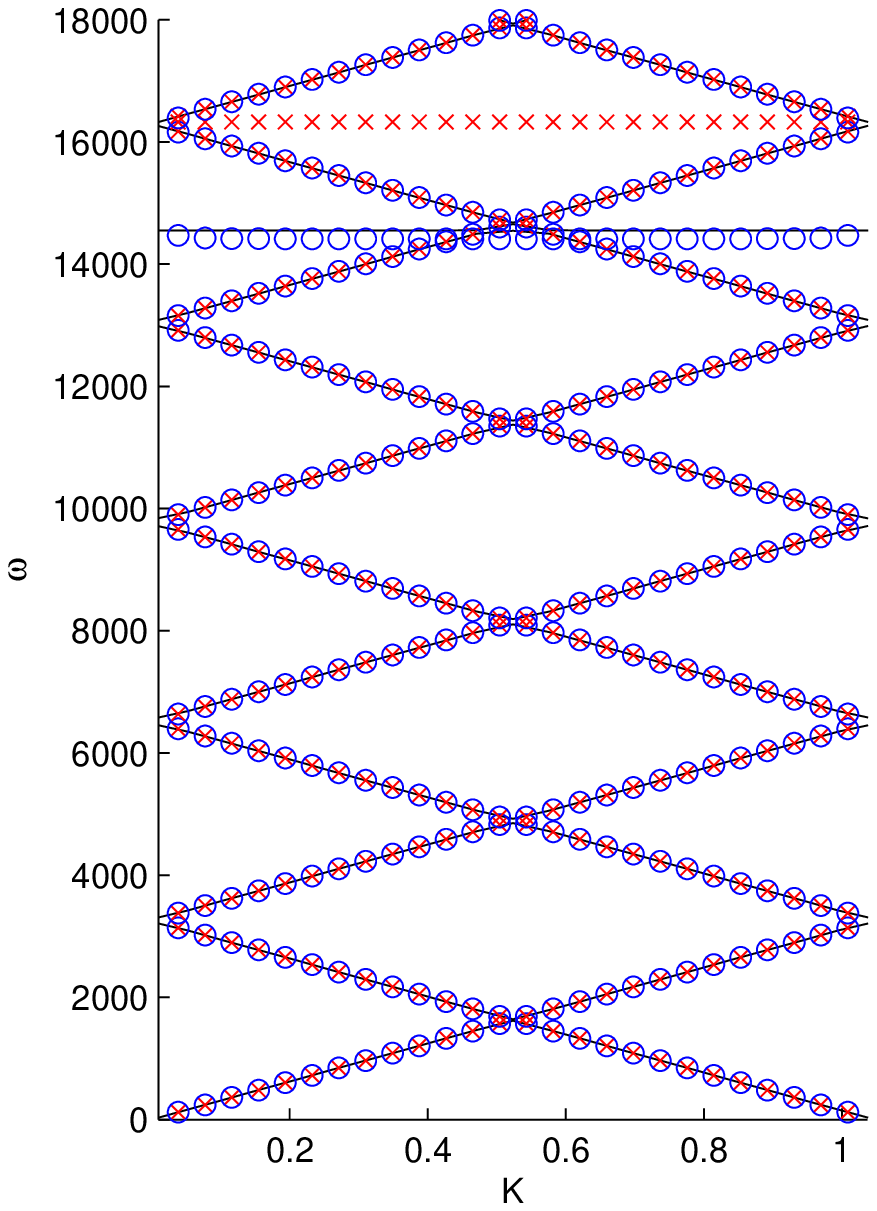}
 \includegraphics[width=0.45\linewidth]{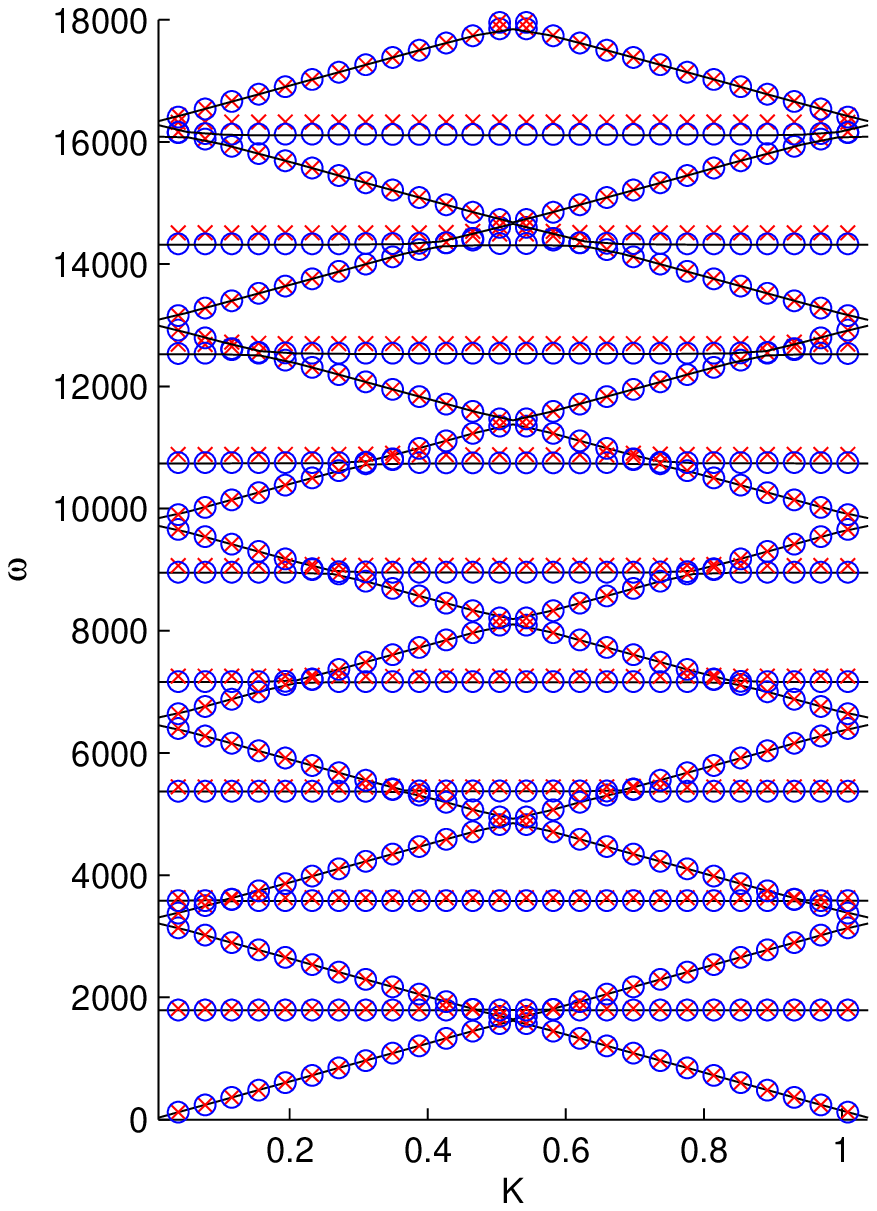}
  \caption{{\footnotesize{Dispersion diagrams for a perfect-interface strip composed of equal thicknesses of aluminium and magnesium {\bf Left:} Short crack ($l=0.6$m) {\bf Right:} Long crack ($l=5.4$m). The solid black lines show the finite element results, while red crosses ({\textcolor{red}{$\times$}}) show the zero order approximation and blue circles (\textcolor{blue}{$\circ$}) show the corrected regime.}}}
  \label{longshort}
\end{figure}
The dispersion diagrams for the cases with short and long crack lengths, again with materials of similar wave speeds, perfect interfaces and the same thicknesses of both material are contained within Figure \ref{longshort}. As one would expect, the length of the crack does not significantly alter the eigenfrequencies of those waves that propagate through the strip (again the correction is on the level of $10^{-4}\%$), since they are not strongly influenced by the presence of the crack. In similar agreement with physical intuition, the first standing wave for the long crack is of much lower frequency than in the geometry housing the particularly short crack. The correction offered by the first order approach is relatively small in the long crack case, but since the zero order model in this case already gave good accuracy with only a 1.3\% discrepancy for the first standing wave, this is not surprising. The corrected eigenfrequency of this wave agrees with finite element results to within $10^{-3}\%$. This can be seen as a surprisingly effective correction since in the long crack geometry, the crack tips are close to the ends of the elementary cell. This gives the boundary layers surrounding the crack tips a small area in which to decay so that they do not influence the Bloch-Floquet conditions.

In the case of the short crack, the zero order approximation of the first standing wave eigenfrequency is easily seen to be significantly different to the true value found in the finite element simulation, with a 12.2\% discrepancy. After applying the correction method, the discrepancy decreases to 0.95\%.

\subsubsection{Materials with more contrasting wave speeds}
\begin{figure}[b]
 \includegraphics[width=0.45\linewidth]{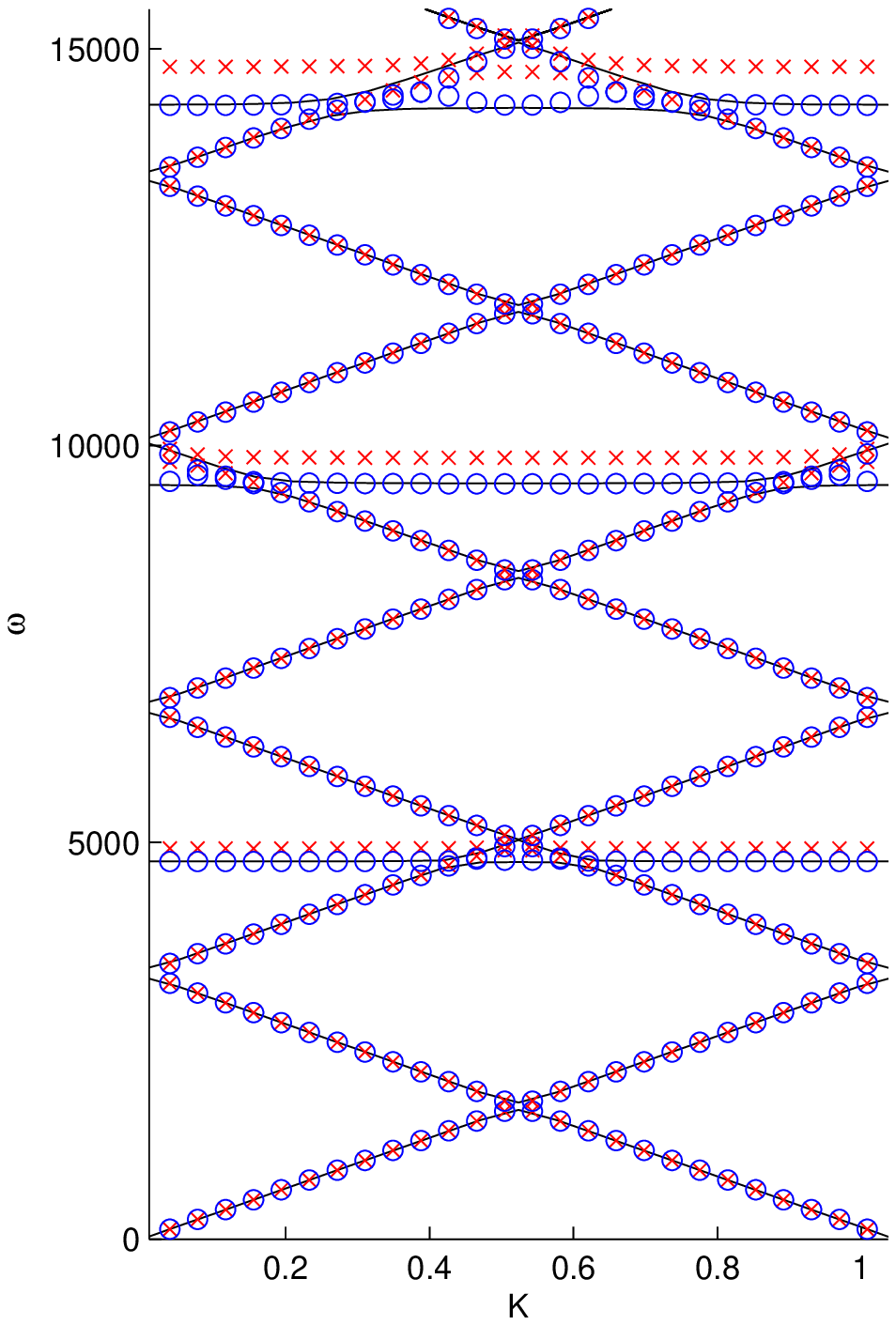}
 \includegraphics[width=0.45\linewidth]{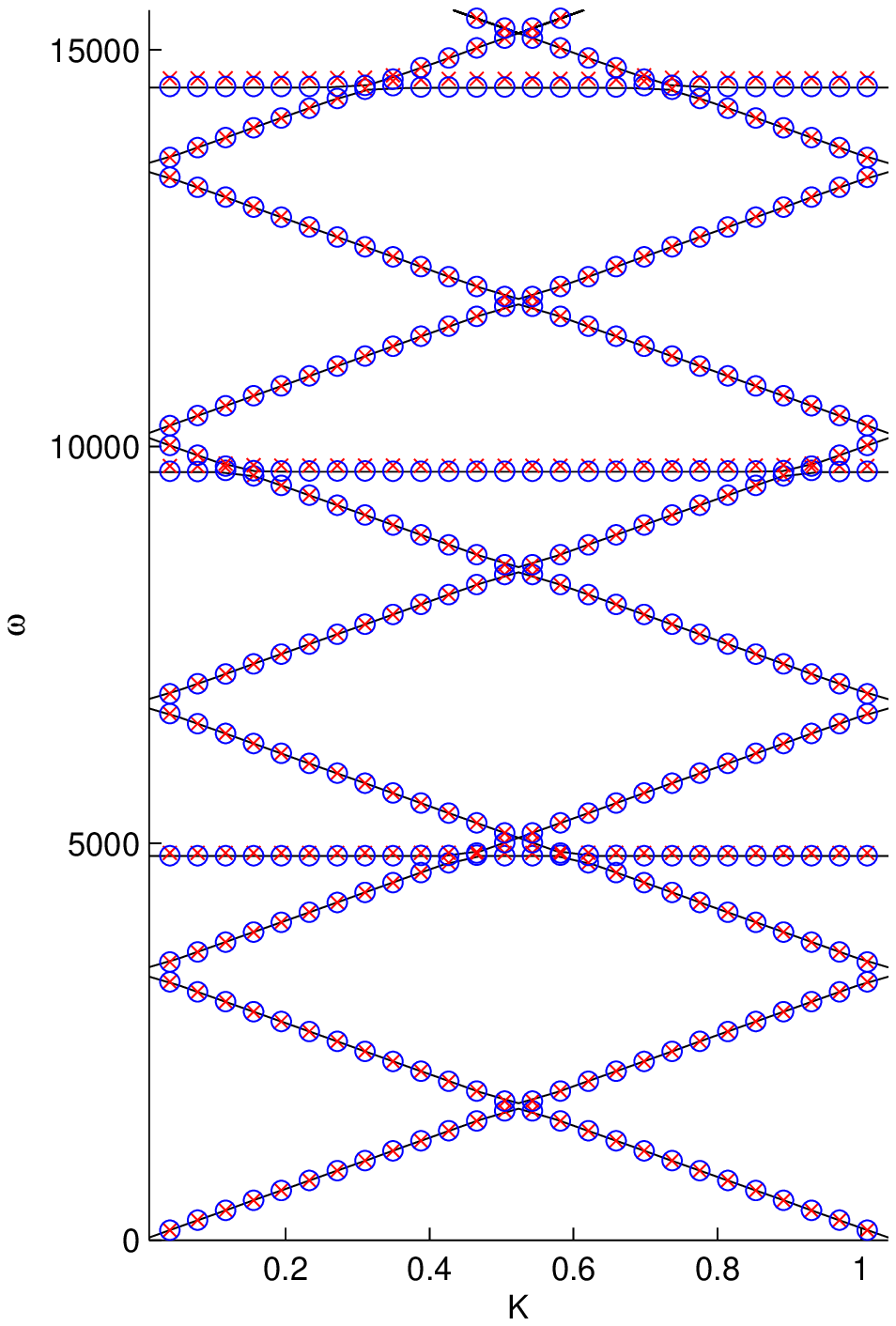}
  \caption{{\footnotesize{Dispersion diagrams for a perfect-interface strip composed of aluminium and iron containing a medium-length crack ($l=2$m). {\bf Left:} Equal thicknesses of iron and aluminium. {\bf Right:} Different thicknesses ($\varepsilon H_1=0.01$m thickness of aluminium, $\varepsilon H_2=0.14$m thickness of iron). The solid black lines show the finite element results, while red crosses ({\textcolor{red}{$\times$}}) show the zero order approximation and blue circles (\textcolor{blue}{$\circ$}) show the corrected regime.}}}
  \label{diffmaterials}
\end{figure}
The standing wave dispersion diagrams for a strip of aluminium and magnesium are presented in Figure \ref{diffmaterials}, for both the symmetrical and asymmetrical cases. The correction is largest, as is the zero order discrepancy, in the symmetrical case. The correction in the case of materials with different wave speeds is less uniform than in the Al-Mg case; this can be readily seen in the highest frequency standing wave shown in the left hand subfigure of Figure \ref{diffmaterials}. The correction still offers a significant improvement in most cases, however, although it is harder to quantify the exact size of a typical discrepancy.

\subsection{Imperfect interface}
\begin{figure}[b]
 \includegraphics[width=0.45\linewidth]{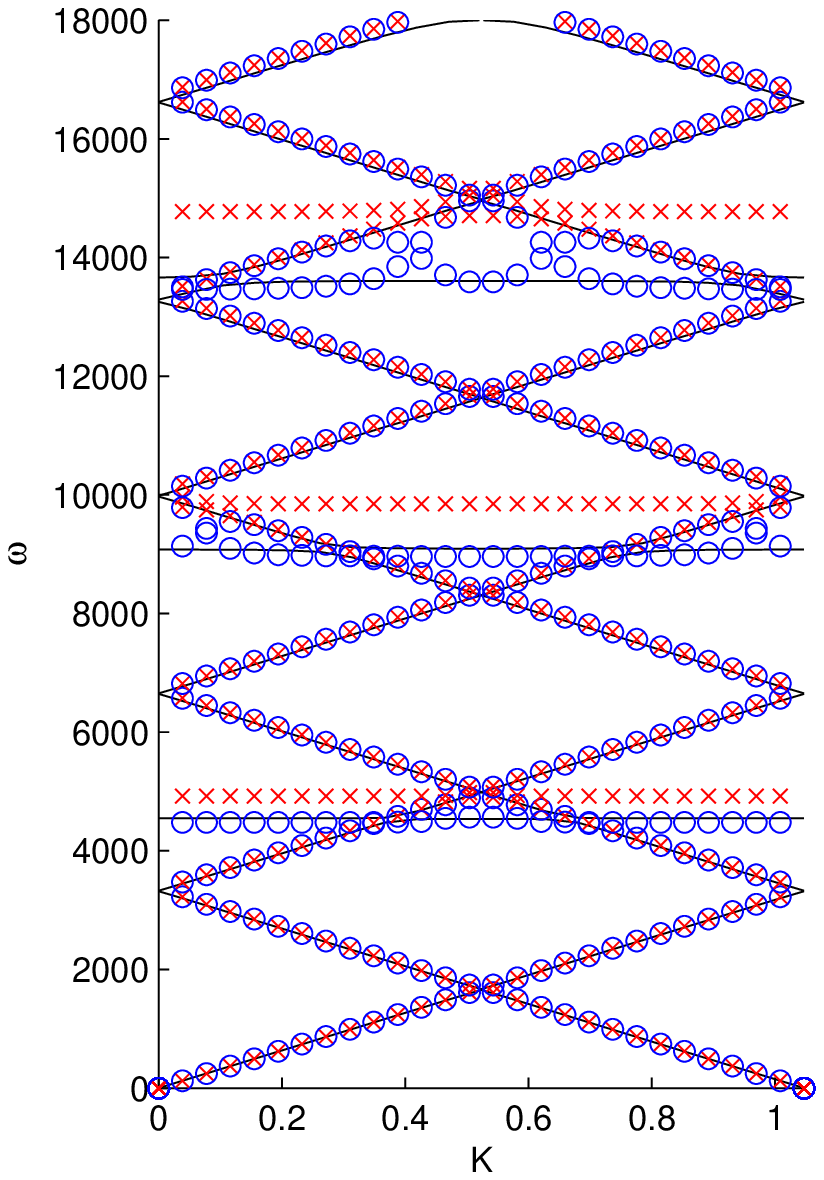}
 \includegraphics[width=0.45\linewidth]{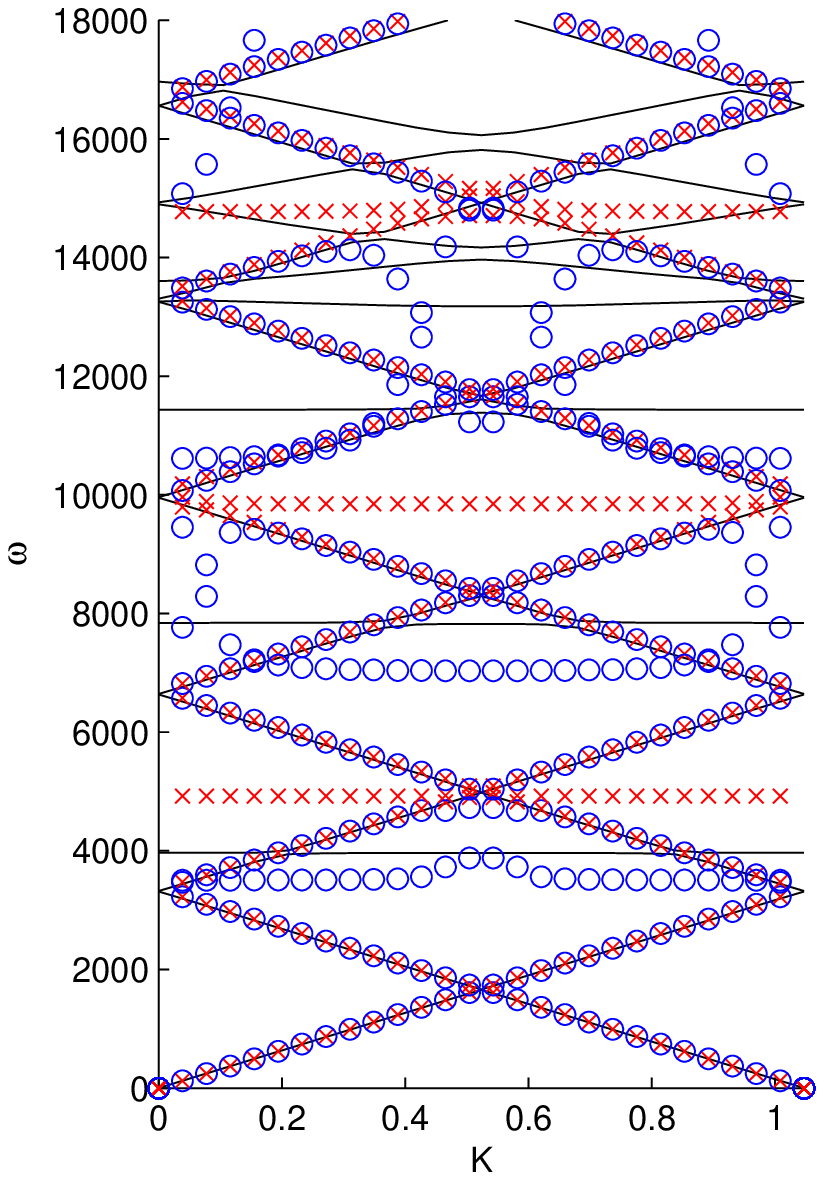}
  \caption{{\footnotesize{Dispersion diagrams for imperfect-interface strips composed of equal thicknesses of aluminium and iron {\bf Left:} Bonding material corresponding to epoxy resin ($\kappa_*=2.88$). {\bf Right:} A highly imperfect interface ($\kappa_*=28.8$) representing an extremely soft bonding material. This rightmost subfigure demonstrates an example in which the low dimensional model is not effective. The solid black lines show the finite element results, while red crosses ({\textcolor{red}{$\times$}}) show the zero order approximation and blue circles (\textcolor{blue}{$\circ$}) show the corrected regime.}}}
  \label{imperfect_figure}
\end{figure}

\begin{figure}[t]
\begin{center}
 \includegraphics[width=0.5\linewidth]{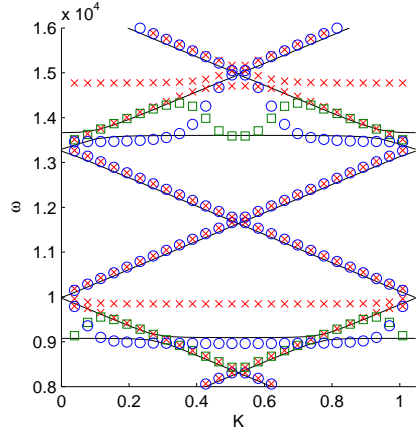}
\end{center}
   \caption{{\footnotesize{Zoomed section of the imperfect interface case, showing how the eigenfrequency correction method causes a crossing-over phenomenon.}}}
  \label{zoomed}
\end{figure}

The results in the case of the imperfect interface analogue follow broadly the same qualitative pattern as in the perfect interface case. The left hand subfigure of Figure \ref{imperfect_figure} gives the dispersion diagram for an iron-aluminium strip, joined with a thin layer of epoxy resin adhesive. Due to the different wavespeeds, the sizes of the standing wave corrections are dependent on the Bloch-Floquet parameter $K$, but in most cases the correction gives a significant improvement in accuracy. An interesting phenomenon can be observed when eigenvalues are close to each other in this subfigure; a zoomed section of the dispersion diagram to illustrate this is given in Figure \ref{zoomed}. In Figure \ref{zoomed}, some of the circles have been replaced by squares; these are the points which approximated propagating waves in the zero-order model which are corrected to approximate the standing waves for some values of $K$. In doing so, a crossing-over phenomenon occurs, where the order of eigenfrequencies switches after correction. The phenomenon becomes more pronounced at higher frequencies.

The right hand figure of Figure \ref{imperfect_figure} corresponds to a case where the materials are bonded in a highly imperfect fashion, using an adhesive with a shear modulus just a tenth that of the epoxy resin whose parameters are given in Appendix \ref{materialconsts} (i.e. a material with shear modulus $2.5\times10^8\textrm{Nm}^{-2}$ and mass density $1850\mathrm{kgm}^{-3}$). This subfigure serves the purpose of presenting a case in which the low dimensional model ceases to provide {\em entirely useful} approximations. Here the finite element simulation displays qualitatively different features which the low dimensional model does not predict at all. The fact that the low dimensional model breaks down is not surprising; in fact, the manuscript \cite{Vellender2011} derives a condition on $\kappa$ for the analysis on which the low dimensional model is based to be valid. The key point of this condition is that if $\kappa$ is too large, the boundary layers $W_A$ and $W_B$ (see equation (\ref{uansatzold}) on page \pageref{uansatzold}) decay sufficiently slowly for the assumption that they are independent to cease to hold. Moreover, if the boundary layers decay slowly from the crack tip, the Bloch-Floquet conditions will be influenced by the boundary layers. This case is interesting in itself and requires separate analysis. This phenomenon has been discussed in \cite{Orlando2003}.

\subsection{Conclusions}
The comparisons between the low dimensional model and the finite element simulations demonstrate that the proposed method of eigenfrequency correction is highly effective in most cases, typically improving accuracy for the standing wave eigenfrequency by an order of magnitude. Moreover, the correction itself is computationally very efficient. The correction becomes less uniform for materials with significantly different wave speeds, improving the accuracy by different amounts in different parts of the dispersion diagram, and misses qualitative features for setups with highly imperfect interfaces. Practically, however, such highly imperfect interfaces are unlikely to be encountered. Fracture parameters are not affected by the analysis and for that reason are omitted in this paper and refer the reader to \cite{Mish2007,Vellender2011}, where discussions and details relating to these parameters can be found. We only underline here that SIF (in the perfect interface case) or COD (in the imperfect interface case) can be constructed as functionals on the low dimensional model without further work. For that reason, the results of this paper are of additional importance.

\appendix

\section{Material constants used for computations in Section \ref{section:numerics}}\label{materialconsts}
\begin{center}
\begin{tabular}{c c c c}
\hline
Material & Shear modulus  & Mass density  & Wave speed \\ 
 & [Nm$^{-2}$] & [kgm$^{-3}$] & [ms$^{-1}$]\\
\hline
Iron & $82 \times10^9$ & 7860 & 3230\\
Magnesium & $17 \times 10^9$ & 1738& 3128\\
Aluminium & $26\times10^9$ &2700 & 3103\\
Epoxy resin & $2.5\times10^9$ & 1850 & 1162\\ \hline
\end{tabular}
\end{center}

\section{Details of parameters varied for computations in Section \ref{section:numerics}}\label{parametersappendix}
\begin{enumerate}
 \item {\bf Type of interface} 
  \begin{itemize}
    \item {\sc Perfect}.
    \item {\sc Imperfect} --- in the finite element computations, a thin layer of epoxy resin is used. This corresponds to a value of $\kappa_*=2.88$ in the asymptotic model.
    \item {\sc Highly imperfect} --- in this case, the bonding material has a shear modulus a tenth that of epoxy resin. This corresponds to a value of $\kappa_*=28.8$.
  \end{itemize}
\item {\bf Length of crack}
  \begin{itemize}
   \item {\sc Short} --- Crack length of $l=a/10=0.6$m.
   \item {\sc Medium} --- Crack length of $l=a/3=2$m. This can be viewed as a `sensible' crack length.
   \item {\sc Long} --- Crack length of $l=9a/10=5.4$m.
  \end{itemize}
\item {\bf Materials}
  \begin{itemize}
   \item {\sc Iron/Aluminium} --- see Appendix \ref{materialconsts} for shear moduli, densities and wave speeds.
   \item {\sc Magnesium/Aluminium} --- both materials have similar wave speeds.
  \end{itemize}
\item {\bf Thicknesses of each material}
  \begin{itemize}
   \item {\sc Symmetrical geometry} --- $\varepsilon H_1=\varepsilon H_2=0.075$m.
   \item {\sc Asymmetrical geometry} --- $\varepsilon H_1=0.01\mathrm{m},$ $\varepsilon H_2=0.14\mathrm{m}$.
  \end{itemize}
\end{enumerate}

\end{document}